\documentclass[a4paper, superscriptaddress]{article}

\usepackage[paperwidth=205mm,paperheight=290mm,top=17mm,bottom=25mm,
inner=17mm,outer=17mm,
twoside]{geometry}
\usepackage{amsmath}
\usepackage{amsfonts}
\usepackage{amssymb}
\usepackage[dvips]{graphicx}
\usepackage{color}
\usepackage[unicode=true,colorlinks=true,linkcolor=magenta, urlcolor=blue, citecolor = blue,breaklinks]{hyperref}
\usepackage{multirow}
\usepackage{url}
\usepackage{breakurl}
\graphicspath{.\\}
\DeclareGraphicsExtensions{.pdf}

\begin{document}

\title{Inflationary Slow-Roll Parameters in the Jordan Frame for Cosmological $F(R)$ Gravity Models} 

\author{Vsevolod~R.~Ivanov$^{a,}$\footnote{E-mail: vsvd.ivanov@gmail.com}\vspace*{3mm}\\
\small  $^a$ Faculty of Physics, Lomonosov Moscow State University,\\
\small Leninskiye Gory~1, 119991, Moscow, Russia}

\date{ \ }

\maketitle

\begin{abstract}
$F(R)$ inflationary models are analyzed without the Weyl transformation to the Einstein frame. Sufficient conditions for existence of global inflationary attractors in $F(R)$ models are provided. Following that, the procedure for calculating inflationary observables completely in the original, Jordan, frame is given. The described procedure is used to analyze a new $F(R)$ inflationary model (a one-parameter deformation of the Starobinsky model), which is shown to provide predictions in good agreement with the recent ACT data.
\end{abstract}


\section{Introduction}\label{intro}

Cosmic inflation remains the leading hypothesis describing the early-stage evolution of the Universe. Its key verifiable predictions are the properties of the Cosmic Microwave Background (CMB), such as spectral tilt index of scalar perturbations $n_s$, tensor-to-scalar power ratio $r$, amplitude of scalar perturbations $A_s$, and scalar perturbations running index $\alpha_s$.

One of the most actively investigated generalizations of the general relativity is the $F(R)$ gravity~\cite{1, 2, 3, 4, 5}.
It so happens that one of the most well-known and long-standing models of inflation is an $F(R)$ model --- the Starobinsky $R + R^2$ model~\cite{6}. Until very recently, it was able to fit the observations well, with the only parameter being the scalaron mass $m$, which is a very appealing feature for any theory, especially considering that the spectral properties of the CMB (such as $n_s$) do not depend on $m$ at all.

However, in light of the most recent Atacama Cosmology Telescope (ACT) data release~\cite{7, 8}, the Starobinsky model is challenged~\cite{9}. This fact motivates search for generalizations of the Starobinsky model, which might conform with the latest observations.
Even disregarding this fact, generalizations of the Starobinsky model are extensively studied in the literature~\cite{10, 11, 12, 13, 14, 15, 16, 17, 18, 19, 20, 21, 22, 23, 24, 25, 26, 27, 28}.

The usual approach for studying cosmological $F(R)$ models (and, more generally, single-field models with nonminimal coupling of the field to gravity) involves the Weyl transformation from the initial frame (usually called the Jordan frame) to the new frame (usually called the Einstein frame), in which the action of the model takes the form of the general relativity action plus the action of a minimally coupled to gravity scalar field. As an example, the action of the Starobinsky model,
\begin{equation*}
S_\mathrm{Star.} = \frac{M_\mathrm{Pl}^2}{2}\int d^4 x \sqrt{-g} \left(R + \frac{R^2}{6 m^2}\right),
\end{equation*}
where $M_\mathrm{Pl}$ is the reduced Planck mass and $m \sim 10^{-5}M_\mathrm{Pl}$ is the inflaton (scalaron) mass (we use the ``$\hbar = c = k_B = 1$'' units and the ``mostly plus'' spacetime signature), has the dual action in the Einstein frame of the form
\begin{equation*}
S^{(E)}_\mathrm{Star.} = \int d^4 x \sqrt{-g^{(E)}} \left(\frac{M_\mathrm{Pl}^2}{2} R^{(E)} - \frac12 {g^{(E)}}^{\mu \nu} \partial_\mu \phi \partial_\nu \phi - V^{(E)}_\mathrm{Star.}(\phi)\right),
\end{equation*}
where
\begin{equation*}
V^{(E)}_\mathrm{Star.}(\phi) = \frac34 M_\mathrm{Pl}^2 m^2 \left(1 - e^{-\sqrt{\frac23}\frac{\phi}{M_\mathrm{Pl}}}\right)^2
\end{equation*}
is the Einstein frame potential of the Starobinsky model, and $\phi$ is the canonical inflaton field. We denote the Einstein frame quantities with the superscript ${}^{(E)}$.

However, there is a methodological interest in studying inflationary $F(R)$ models (and the models with nonminimal coupling in general) in the initial, Jordan, frame~\cite{29, 30, 31, 32}. Firstly, it is not often possible to find explicit formulae which connect the dynamical quantities of the different frames. Secondly, the question arises as to which of the two frames should be considered ``physical''. Thirdly, for some models, it is entirely possible that different expressions of interest (for example, the slow-roll parameters) in the Jordan frame look much simpler than the corresponding ones in the Einstein frame, and, therefore, are easier to study and generalize.

The goal of this work is twofold. The first goal is a methodological one: to describe a procedure for calculating inflationary parameters without need to resort to the Einstein frame description. The second goal is to find a simple, one-parameter deformation of the Starobinsky model which conforms with the latest observational data.

\section{$F(R)$ model of gravity. Equations of the model}

\subsection{Action of the model. The equivalent action}

The $F(R)$ gravity model is described by the action
\begin{equation}
S_F = \int d^4 x \sqrt{-g} F(R),
\end{equation}
where $F(R)$ is an arbitrary function of the Ricci curvature $R$.
Now we show that, for reasonable $F(R)$ functions, it is possible to use the equivalent action
\begin{equation}
S_J = \int d^4 x \sqrt{-g} \left(U(\sigma) R - V(\sigma)\right),
\end{equation}
where
\begin{equation}
U(\sigma) = F_{,\sigma}(\sigma),\quad V(\sigma) = \sigma U(\sigma) - F(\sigma).
\end{equation}
By varying the action $S_J$ with respect to the auxiliary field $\sigma$, one obtains the equation
\begin{equation}
U_{,\sigma}(\sigma)(R - \sigma) = 0,
\end{equation}
from which it follows that, if $U_{,\sigma}(\sigma) \not\equiv 0$, then $R = \sigma$.
This equality, in turn, ensures equivalence of the actions $S_F$ and $S_J$.
From now on, we work with the action $S_J$ instead of $S_F$.

\subsection{Equations of the model in general form}

The metric equations of the model in question are
\begin{equation}
\label{s_j_metric_eqn}
U(\sigma)\left(R_{\mu \nu} - \frac12 R g_{\mu \nu}\right) = \nabla_\mu \nabla_\nu U - g_{\mu \nu} \Box U - \frac12 V(\sigma) g_{\mu \nu},
\end{equation}
and the field equation is
\begin{equation}
\label{s_j_field_eqn}
R U_{, \sigma} - V_{, \sigma} = 0 \implies R = V_{, \sigma} / U_{, \sigma} = V_{, U}.
\end{equation}
We also write the trace equation (it is obtained by contracting~(\ref{s_j_metric_eqn}) with $g^{\mu \nu}$)
\begin{equation}
-U R = -3 \Box U - 2 V.
\end{equation}
By combining it with the field equation, one obtains the equation
\begin{equation}
\label{s_j_u_eqn}
\Box U - \frac13 \left(V_{, U} U - 2 V\right) = \Box U - W_{, U} = 0,
\end{equation}
where we have introduced the effective potential
\begin{equation}
 W(U) = \frac13\int^{U}_{U_0} dU' \left(V_{, U'}(U')U' - 2 V(U')\right).
\end{equation}
Here $U_0 = \left. U\right|_{\sigma = 0}$.

\subsection{Equations of the model in spatially flat FLRW metric}

Now we write the equations of our model in the spatially flat Friedmann-Lema\^\i tre-Robertson-Walker (FLRW) metric, with its line element being described by
\begin{equation}
ds^2  = -dt^2 + a^2(t) \left(dx^2 + dy^2 + dz^2\right).
\end{equation}
In this metric, Eqs.~(\ref{s_j_metric_eqn}) reduce to
\begin{equation}
\label{flrw_metric_eqn}
\begin{aligned}
3 U H^2 &= -3 H \dot{U} + \frac12 V,\\
-U\, \left(2 \dot{H} + 3 H^2\right) &= \ddot{U} + 2 H \dot{U} - \frac12 V,
\end{aligned}
\end{equation}
and field equation~(\ref{s_j_field_eqn}) reduces to
\begin{equation}
\label{flrw_field_eqn}
6 \left(\dot{H} + 2 H^2\right) U_{, \sigma} - V_{, \sigma} = 0.
\end{equation}
Here, ``dots'' represent $d / dt$, and $H = \dot{a}/a$ is the Hubble parameter.

Eq.~(\ref{s_j_u_eqn}) in the FLRW metric reduces to
\begin{equation}
\label{flrw_u_eqn}
\ddot{U} + 3 H \dot{U} + W_{, U} = 0.
\end{equation}

\section{Stability analysis of the model}

Now we check whether the model in question can describe slow-roll inflation. One of the key features of standard slowly-rolling inflationary models is the existence of an inflationary attractor. Moreover, if the model has an asymptotically stable point at $U = M_\mathrm{Pl}^2/2$ and $V = 0$, then it has a graceful exit from the inflationary stage. In other words, at dynamical equilibrium, the $S_J$ action is equal to the standard Einstein-Hilbert action of general relativity, and all possible trajectories tend to this equilibrium state.

We start with rewriting Eq.~(\ref{flrw_u_eqn}) as a dynamical system with its equilibrium point being $(0, 0)$. Firstly, we need to express $H$ as a function of $U$ and $\dot{U}$. Using the first equation of~(\ref{flrw_metric_eqn}), one obtains
\begin{equation}
H(U, \dot{U}) = \sqrt{\left(\frac{\dot{U}}{2 U}\right)^2 + \frac{V(U)}{6 U}} - \frac{\dot{U}}{2 U}.
\end{equation}
Note that we take only the solution corresponding to nonnegative $H$, since this particular solution is the one relevant to the inflation. We emphasize that $H$ is \textit{always} nonnegative for positive $U$, provided $V$ is nonnegative as well. We also assume that $V$ possesses no singular points in the domain of interest.

Then, we introduce
\begin{equation}
q = \frac{2}{M_\mathrm{Pl}^2}U - 1 = u - 1, \quad p = \dot{q} = \dot{u}, \quad v = \frac{2}{M_\mathrm{Pl}^2}V,  \quad w_{, q} = \frac{2 W_{, q}}{M_\mathrm{Pl}^2}=\frac13\left(v_{, q}(q+1) - 2v\right),
\end{equation}
and rewrite Eq.~(\ref{flrw_u_eqn}) as a dynamical system:
\begin{equation}
\label{dyn_syst}
\left\{
\begin{aligned}
\dot{q} &= p,\\
\dot{p} &= -3 H(q, p) p - w_{, q}(q).
\end{aligned}
\right.
\end{equation}
We see that $(q, p) = (0, 0)$ is, indeed, a stable point of the system, provided $w_{,q}(0) = 0$.

Let's introduce the Lyapunov candidate function
\begin{equation}
V_\mathrm{Lyap.} = \frac{p^2}{2} + w(q).
\end{equation}
Its time derivative due to system~(\ref{dyn_syst}) reads
\begin{equation}
\label{lyap_dt}
\dot{V}_\mathrm{Lyap.} = \frac{\partial V_\mathrm{Lyap.}}{\partial q} \dot{q} + \frac{\partial V_\mathrm{Lyap.}}{\partial p} \dot{p} = -3 H p^2.
\end{equation}

Now we use Lyapunov's second method for stability: if $V_\mathrm{Lyap.}$ is such that
\begin{equation}
\begin{aligned}
V_\mathrm{Lyap.} &= 0 \iff (q, p) = (0, 0),\\
V_\mathrm{Lyap.} &> 0 \iff (q, p) \neq (0, 0),\\
\dot{V}_\mathrm{Lyap.} &\leq 0,
\end{aligned}
\end{equation}
then $V_\mathrm{Lyap.}$ is the Lyapunov function of dynamical system~(\ref{dyn_syst}), and $(q, p) = (0, 0)$ is a stable point of the system. We see that these criteria are satisfied if $w(q) > 0$ for $q \neq 0$, and $w(q) = 0$ for $q = 0$.

For our purposes, however, simply having stability is not enough; we need \textit{asymptotic} stability. Lyapunov's second method ensures asymptotic stability only if the $\dot{V}_\mathrm{Lyap.}$ is negative definite for all $(q, p) \neq (0, 0)$, which is not the case for our choice of $V_\mathrm{Lyap.}$.

But not all hope is lost: due to the fact that our system is autonomous, we can prove asymptotic stability of the $(0, 0)$ equilibrium point using the Barbashin-Krasovskii-LaSalle principle~\cite{33, 34}, even for negative \textit{semi}definite $\dot{V}_\mathrm{Lyap.}$. According to this principle, if we can show that the only complete trajectory of the dynamical system, which satisfies the condition $\dot{V}_\mathrm{Lyap.} = 0$ for all $t$, is the trivial solution $(q(t), p(t)) = (0, 0)$, then the trivial solution is asymptotically stable (moreover, if $w(q)$ is unbounded from above, then the trivial solution is globally asymptotically stable). Let's see that this condition is satisfied under certain requirements for $w$ and $v$.

Looking at~(\ref{lyap_dt}), one can see that $\dot{V}_\mathrm{Lyap.} = 0$ in two cases (not counting the trivial case): $p = 0$ and $H = 0$.

We analyze the ``$p = 0, \, q \neq 0$'' case first. System~(\ref{dyn_syst}) in this case becomes
\begin{equation*}
\left\{
\begin{aligned}
\dot{q} &= 0,\\
\dot{p} &= -w_{, q}(q).
\end{aligned}
\right.
\end{equation*}
From here it is obvious that any $(q, p)$ point besides $(0, 0)$ will leave the set given by $\dot{V}_\mathrm{Lyap.} = 0$, provided that $w_{, q}(q) \neq 0$ for $q \neq 0$.

In the case ``$H = 0$'', system~(\ref{dyn_syst}) becomes
\begin{equation*}
\left\{
\begin{aligned}
\dot{q} &= p,\\
\dot{p} &= -w_{, q}(q).
\end{aligned}
\right.
\end{equation*}
From this result follows the same conclusion: any $(q, p)$ point besides $(0, 0)$ will leave the set given by $\dot{V}_\mathrm{Lyap.} = 0$, provided that $w_{, q}(q) \neq 0$ for $q \neq 0$.

The analysis given above shows that the trivial solution is asymptotically stable, and we have a graceful exit from inflation, provided
\begin{equation}
\label{attr_cond}
\begin{aligned}
w(q) &> 0 \iff q \neq 0,\\
w_{, q}(q) &= 0 \iff q = 0,\\
v(q) &> 0 \iff q \neq 0,\\
v(q) &= 0 \iff q = 0.
\end{aligned}
\end{equation}
The condition on $w_{, q}$ probably can be slightly weakened, in principle; it seems reasonable to assume that our conclusions about stability would still hold if $w_{, q}(q) = 0 \iff w_{, qq}(q) = 0$, but we do not investigate this case in this paper.

\section{Slow-roll parameters in the Jordan frame}

\subsection{Defining the slow-roll parameters}
In contrast with the usual, Einstein frame-based approach, one needs two sets of slow-roll parameters in the Jordan frame. The first set is
\begin{equation}
\label{sr_h_def}
\epsilon_H = -\frac{\dot{H}}{H^2}, \, \eta_H = \frac{\dot{\epsilon}_H}{H \epsilon_H}, \, \cdots,
\end{equation}
and the second one is
\begin{equation}
\label{sr_u_def}
\epsilon_U = -\frac{\dot{U}}{2 H U}, \, \eta_U = \frac{\dot{\epsilon}_U}{H \epsilon_U}, \, \cdots.
\end{equation}
Now we rewrite Eqs.~(\ref{flrw_metric_eqn}) in terms of these parameters:
\begin{equation}
\label{sr_metric_eqn}
\begin{aligned}
3 U H^2\left(1 - 2 \epsilon_U\right) &= \frac12 V,\\
-U H^2\, \left(3 - 2\epsilon_H - 4 \epsilon_U\right) &= 2 U H^2 \left(\epsilon_H + 2 \epsilon_U - \eta_U\right) - \frac12 V.
\end{aligned}
\end{equation}
After adding the first equation to the second one and dividing both sides by a common factor, one obtains the useful relation
\begin{equation}
\epsilon_H - \epsilon_U = \epsilon_U \left(\epsilon_H + 2 \epsilon_U - \eta_U\right).
\end{equation}
This relation allows us to present $\epsilon_H$ completely in terms of $\epsilon_U$ and $\eta_U$:
\begin{equation}
\label{eps_H_of_eps_U_eta_U}
\epsilon_H = \epsilon_U \frac{1 + 2 \epsilon_U - \eta_U}{1 - \epsilon_U}.
\end{equation}

\subsection{Slow-roll parameters under the attractor approximation}
Let's assume that conditions~(\ref{attr_cond}) are satisfied, and the initial conditions for $U$ and $\dot{U}$ are sufficiently far away from the equilibrium point. The attractor approximation comes down to neglecting the $\ddot{U}$ term in Eq.~(\ref{flrw_u_eqn}). Thus, we have
\begin{equation*}
3 H \dot{U} \approx -\frac13 \left(V_{,U} U - 2 V\right).
\end{equation*}
In terms of the slow-roll parameters, this result reads
\begin{equation*}
\epsilon_U \approx \frac{V_{,U} U - 2 V}{3 V}\left(1 - 2 \epsilon_U\right).
\end{equation*}
Solving for $\epsilon_U$, we finally obtain
\begin{equation}
\label{eps_u_psr}
\epsilon_U \approx \frac{V_{, U} U - 2 V}{2 V_{, U} U - V} = \frac{V_{, \sigma} U - 2 U_{, \sigma} V}{2 V_{, \sigma} U - U_{, \sigma} V} \equiv \hat{\epsilon}_U.
\end{equation}
So, we have obtained a result which resembles the traditional Potential Slow-Roll (PSR) parameters, but for $F(R)$ gravity models, in the Jordan frame.

Having the result given by Eq.~(\ref{eps_u_psr}), we can easily obtain the same, PSR-like, formulae for $\eta_U$ and $\epsilon_H$. For $\eta_U$, we have the exact relation
\begin{equation}
\eta_U = \frac{\dot{\epsilon}_U}{H \epsilon_U} = \frac{{\epsilon_U}_{, U} \dot{U}}{H \epsilon_U} = -2 U {\epsilon_U}_{, U} = - 2\frac{U}{U_{, \sigma}} {\epsilon_U}_{, \sigma},
\end{equation}
into which we can directly plug~(\ref{eps_u_psr}), and obtain
\begin{equation}
\eta_U \approx -2 U {{\hat{\epsilon}}_{U, U}} \equiv {\hat{\eta}}_U.
\end{equation}
Having now both $\hat{\epsilon}_U$ and $\hat{\eta}_U$, we can obtain an approximation for $\epsilon_H$ in the same, PSR-like, form by using Eq.~(\ref{eps_H_of_eps_U_eta_U}):
\begin{equation}
\epsilon_H \approx \hat{\epsilon}_U \frac{1 + 2 \hat{\epsilon}_U - \hat{\eta}_U}{1 - \hat{\epsilon}_U} \equiv \hat{\epsilon}_H.
\end{equation}
 We do not present the explicit formulae for $\hat{\eta}_U$ and $\hat{\epsilon}_H$ in terms of $U$ and $V$; they are cumbersome, but are very straightforward to obtain.

\subsection{Number of e-folds $N$ in terms of the slow-roll parameters}

The standard unit of time in the context of inflationary cosmology is the number of e-folds $N$, defined as
\begin{equation}
N(t) = \ln \frac{a_{end}}{a(t)},
\end{equation}
where $a_{end}$ is the value of the scale factor $a$ at the end of inflationary stage. The sign of $N$ is chosen in such a way that $N=0$ corresponds to the end of inflation, and positive values of $N$ correspond to some moments of time during (or, possibly, before) inflation.

To express $N$ as a function of $U$ (or $\sigma$), we firstly observe that
\begin{equation*}
\epsilon_U = -\frac{\dot{U}}{2 H U} = \frac{1}{2 U} \frac{d U}{d N},
\end{equation*}
since $dN = -H dt$.
From here it follows that
\begin{equation}
N(U_{in}) = \frac12 \int_{U_{end}}^{U_{in}} \frac{dU}{U \epsilon_U},\quad N(\sigma_{in}) = \frac12 \int_{\sigma_{end}}^{\sigma_{in}} \frac{U_{, \sigma} d\sigma}{U \epsilon_U}.
\end{equation}
Note that these results are exact. The approximate results for $N$ are given by
\begin{equation}
N(U_{in}) \approx \frac12 \int_{U_{end}}^{U_{in}} \frac{dU}{U \hat{\epsilon}_U},\quad N(\sigma_{in}) \approx \frac12 \int_{\sigma_{end}}^{\sigma_{in}} \frac{U_{, \sigma} d\sigma}{U \hat{\epsilon}_U}.
\end{equation}

\subsection{Connection with the Einstein frame PSR parameters}
The usual Einstein frame PSR parameters are
\begin{equation}
\epsilon_V = \frac{M_\mathrm{Pl}^2}{2}\left(\frac{V^{(E)}_{, \phi}}{V^{(E)}}\right)^2, \quad \eta_V = M_\mathrm{Pl}^2\frac{V^{(E)}_{, \phi \phi}}{V^{(E)}}.
\end{equation}
Here
\begin{equation}
V^{(E)} = \frac{M_\mathrm{Pl}^4}{4} \frac{V}{U^2}, \quad \phi = \sqrt{\frac32} M_\mathrm{Pl} \ln \frac{2 U}{M_\mathrm{Pl}^2}.
\end{equation}
After some straightforward calculations, one obtains
\begin{equation}
\epsilon_V = \frac13 \left(\frac{V_{, U}U}{V} - 2\right) = 3{\hat{\epsilon}}_U^2 + o({\hat{\epsilon}}_U^2),
\end{equation}
and
\begin{equation}
\eta_V = 2 \epsilon_U + \sqrt{2}M_\mathrm{Pl}\frac{d \sqrt{\epsilon_V}}{d \phi} = 2 \epsilon_V + 2 U \hat{\epsilon}_{U, U} + o( \hat{\epsilon}_{U}) = -\hat{\eta}_U + o( \hat{\epsilon}_{U}).
\end{equation}

From here it is easy to obtain the leading-order results for the inflationary observables $n_s$, $r$, and $A_s$:
\begin{eqnarray}
n_s &\approx& 1 - 6 \epsilon_V + 2 \eta_V \approx 1 - 2 \hat{\eta}_U,\\
r &\approx& 16 \epsilon_V \approx 48 \hat{\epsilon}_U^2,\\
A_s &\approx& \frac{2 V^{(E)}}{3 \pi^2 M^4_\mathrm{Pl} r} \approx \frac{V}{6 \pi^2 U^2 r}.
\end{eqnarray}

It is interesting to note that the results obtained do not depend on $\hat{\epsilon}_H$. Its only purpose is to indicate the end of inflationary stage, where it should be $\approx 1$.

For the amount of e-folds in the Einstein frame, $N^{(E)}$, we have
\begin{equation}
N^{(E)} \approx \frac{1}{M_\mathrm{Pl}}\int_{\phi_{end}}^{\phi_{in}} \frac{d \phi}{\sqrt{2 \epsilon_V}} \approx \frac12 \int_{U_{end}}^{U_{in}} \frac{dU}{U \hat{\epsilon}_U} \approx N.
\end{equation}
Strictly speaking, the endings of inflation in the Jordan and the Einstein frame need not coincide; however, the difference between $N^{(E)}$ and $N$ is negligible enough for our purposes (it is $O(1)$ at most), and so we assume that $N^{(E)} \approx N$ in what follows.

We also present the formula for calculating the running index $\alpha_s = d \ln n_s / d \ln k$:
\begin{equation}
\alpha_s \approx -\frac{d n_s}{d N^{(E)}} \approx -\frac{d n_s}{d N} \approx 2\frac{d \hat{\eta}_U}{d N} \approx 4 U \hat{\epsilon}_U \frac{d \hat{\eta}_U}{d U}.
\end{equation}

We emphasize that all the results presented in this subsection are valid only in the slow-roll approximation.

\section{Application of the results}

\subsection{The Starobinsky model}
As a ``sanity check'', we apply our results to the Starobinsky $R^2$ model~\cite{6}, with its $F$-function being
\begin{equation}
F(\sigma) = \frac{M_\mathrm{Pl}^2}{2}\left(\sigma + \frac{\sigma^2}{6 m^2}\right).
\end{equation}
The corresponding $U$ and potential $V$ read
\begin{equation}
U(\sigma) =  \frac{M_\mathrm{Pl}^2}{2}\left(1 + \frac{\sigma}{3 m^2}\right) = \frac{M_\mathrm{Pl}^2}{2} u, \quad V(\sigma) =  \frac{M_\mathrm{Pl}^2}{2} \frac{\sigma^2}{6 m^2} = \frac{M_\mathrm{Pl}^2}{2} v.
\end{equation}
It is easy to express $v$ in terms of $u$:
\begin{equation}
v(u) = \frac32 m^2 \left(u - 1\right)^2 \rightarrow v(q) = \frac32 m^2 q^2,\quad q = u - 1.
\end{equation}
The effective potential $w(q)$ is given by
\begin{equation}
w(q) = \frac13 \int^q_0 dq' \left(v_{, q'}(q')(q'+1) - 2v(q')\right) = \frac12 m^2 q^2.
\end{equation}

Now it is obvious that all conditions~(\ref{attr_cond}) are satisfied for this model, and so there exists a unique inflationary attractor. Moreover, the attractor is global, since $w(q)$ is unbounded from above.

Now we calculate the slow-roll parameters in our PSR-like approximation. We have
\begin{equation}
\hat{\epsilon}_U = \frac{2}{3 u + 1} \approx \frac{2}{3 u}, \quad \hat{\eta}_U = \frac{12 u}{\left(3 u + 1\right)^2} \approx \frac{4}{3 u}.
\end{equation}
The expressions following the ``$\approx$'' signs are obtained under the assumption $u \gg 1$.

For the amount of e-folds, we obtain
\begin{equation}
N \approx \frac14 \int_{u_{end}}^{u_{in}} du \left(3 + \frac{1}{u}\right) = \frac34\left(u_{in} - u_{end}\right) + \frac14 \ln \frac{u_{in}}{u_{end}} \approx \frac34 u_{in},
\end{equation}
and so
\begin{equation}
u \approx \frac43 N \implies \hat{\epsilon}_U \approx \frac{1}{2 N}, \quad \hat{\eta}_U \approx \frac{1}{N}.
\end{equation}

Finally, we obtain the inflationary observables:
\begin{equation}
n_s \approx 1 - \frac{2}{N}, \quad r \approx \frac{12}{N^2}, \quad A_s \approx \frac{m^2}{2 \pi^2 M_\mathrm{Pl}^2 r} \approx \frac{m^2 N^2}{24 \pi^2 M_\mathrm{Pl}^2}, \quad \alpha_s \approx -\frac{2}{N^2}.
\end{equation}
These results coincide with the well-known leading-order predictions of the Starobinsky model~\cite{9, 35, 36, 37}, as expected.

\subsection{New model}

Now we present a new model, which is a one-parameter deformation of the Starobinsky model. Its $F$-function is given by
\begin{equation}
F(\sigma) = \frac{M_\mathrm{Pl}^2}{2}\left(\sigma + \frac{\sigma}{3 \delta} \ln \left(1 + \delta \frac{\sigma}{m^2}\right) + \frac{m^2}{3 \delta^2} \left(\ln \left(1 + \delta \frac{\sigma}{m^2}\right) - \delta \frac{\sigma}{m^2}\right)\right).
\end{equation}
where $\delta$ is a new positive dimensionless parameter. The Starobinsky model is recovered in the limit $\delta \rightarrow 0$. This model might seem complicated at a glance, but it has a pretty simple $U$-function,
\begin{equation}
U(\sigma) = \frac{M_\mathrm{Pl}^2}{2}\left(1 + \frac{1}{3 \delta} \ln \left(1 + \delta \frac{\sigma}{m^2}\right)\right) = \frac{M_\mathrm{Pl}^2}{2} u(\sigma),
\end{equation}
which is easily invertible:
\begin{equation}
\sigma(u) = \frac{m^2}{\delta}\left(e^{3\delta(u - 1)} - 1\right) \rightarrow \sigma(q) = \frac{m^2}{\delta}\left(e^{3\delta q} - 1\right).
\end{equation}

\begin{figure}[h!]
\centering
\includegraphics[width=0.9\linewidth]{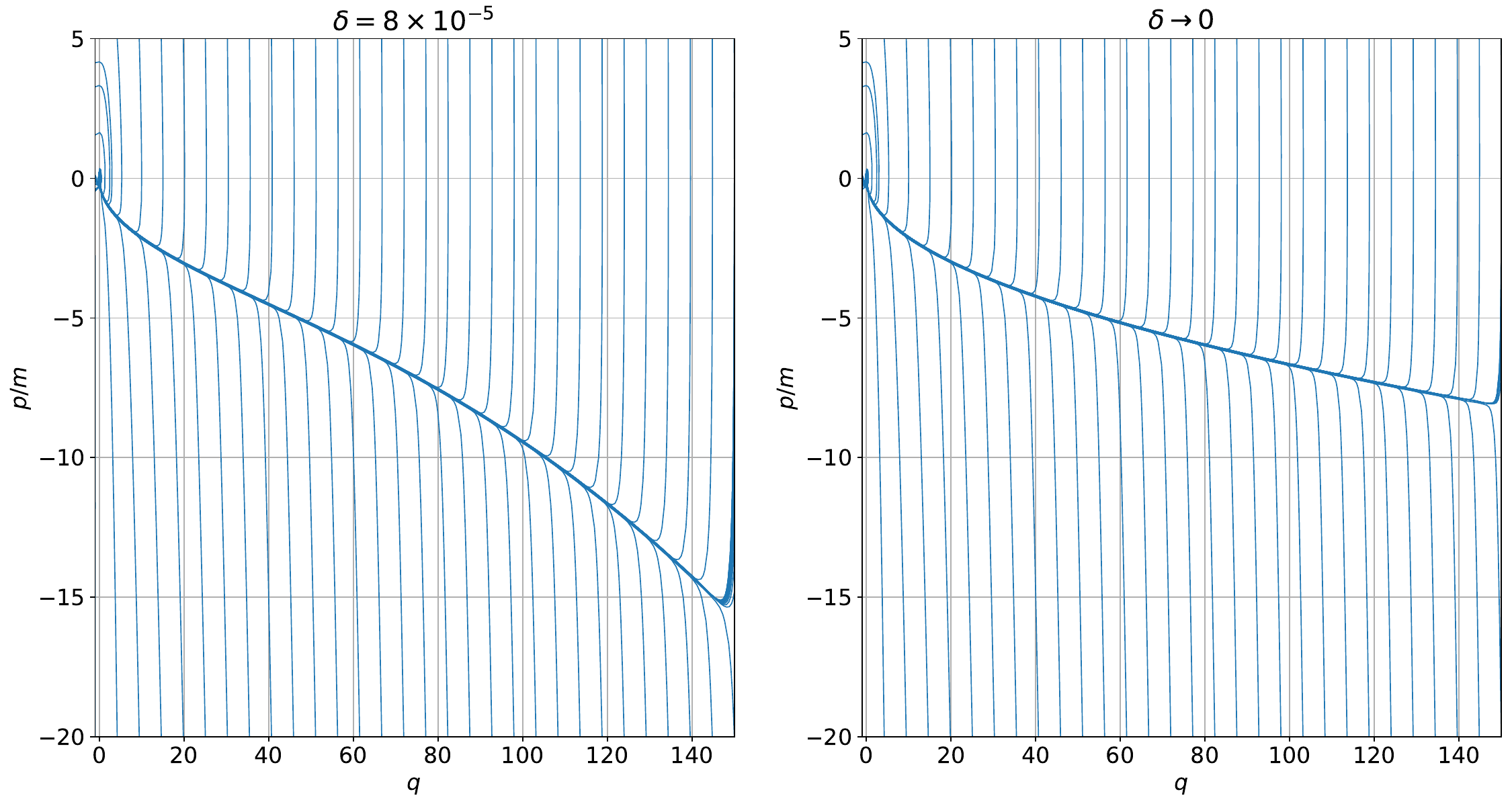}
\caption{Phase portrait of dynamical system~(\ref{dyn_syst}) for the new model (left) in comparison with the phase portrait for the Starobinsky model (right). Inflation ends at $q_{end} \approx 0.3$ in both cases. The values of $q_{in}$ at the horizon crossing (we take $N_{in} = 60$) are different, though; for the new model $q_{in} \approx 87$, and for the Starobinsky model $q_{in} \approx 80$.}\label{fig:phase_portrait}
\end{figure}

The potential $v(q)$ for this model reads
\begin{equation}
v(q) = \frac{m^2}{3 \delta^2}\left(e^{3 \delta q} - 1 - 3 \delta q\right),
\end{equation}
and the effective potential $w(q)$ is given by
\begin{equation}
w(q) = \frac{m^2}{9 \delta^3} \left(\delta (q + 1)\left(e^{3 \delta q} - 1 - 3 \delta q\right) - \left(e^{3 \delta q} - 1 - 3 \delta q - \frac12 (3 \delta q)^2\right)\right).
\end{equation}
It can be shown that these $v$ and $w$ satisfy conditions~(\ref{attr_cond}), and so a unique inflationary attractor does exist. Moreover, it is global. The existence of the attractor can also be clearly seen on a phase portrait of the dynamical system (see Fig.~\ref{fig:phase_portrait}).

For this model, it is also straightforward to write the Einstein frame potential as a function of the canonical inflaton field $\phi$:
\begin{equation}
V^{(E)} = \frac{M_\mathrm{Pl}^4}{4}\frac{V}{U^2} = \frac{M_\mathrm{Pl}^2}{2}\frac{v}{u^2} = \frac{M_\mathrm{Pl}^2 m^2}{6 \delta^2}e^{-2\sqrt{\frac23}\frac{\phi}{M_\mathrm{Pl}}}\left(e^{3 \delta (e^{\sqrt{\frac23}\frac{\phi}{M_\mathrm{Pl}}}-1)} - 1 - 3 \delta (e^{\sqrt{\frac23}\frac{\phi}{M_\mathrm{Pl}}}-1)\right).
\end{equation}

We now derive leading-order predictions for the new model. To achieve that, let's assume that $3 \delta u \sim 1/u \ll 1$. Then we have for $v(u)$
\begin{equation}
v(u) = \frac32 m^2 (u - 1)^2\left(1 + \delta (u - 1)\right) + o\left((\delta u)^3\right).
\end{equation}
The appoximate slow-roll parameter $\hat{\epsilon}_U$ in this case is
\begin{equation}
\label{psr_new_model}
\hat{\epsilon}_U = \frac{2}{3u}\left(1 + \frac12 \delta u^2\right) + o(\delta u),
\end{equation}
and the number of e-folds $N$ can be approximated as
\begin{equation}
N \approx \frac34 \int_{u_{end}}^{u_{in}}\frac{du}{1 + \frac12 \delta u^2} \approx \frac34 \sqrt{\frac{2}{\delta}} \arctan \left(\sqrt{\frac{\delta}{2}}u_{in}\right) \implies u(N) \approx \sqrt{\frac{2}{\delta}} \tan \left(\sqrt{\frac{\delta}{2}}\frac43 N\right).
\end{equation}
Substituting the last result into~(\ref{psr_new_model}), one obtains
\begin{equation}
\hat{\epsilon}_U \approx \frac{1}{2 N} \frac{\sqrt{2 \delta}\frac43 N}{\sin \left(\sqrt{2 \delta} \frac43 N\right)} = \frac{g_\epsilon(N; \delta)}{2 N} \implies \hat{\eta}_U \approx -\frac{d}{d N} \ln \hat{\epsilon}_U = \frac{1}{N} \frac{\sqrt{2 \delta}\frac43 N}{\tan \left(\sqrt{2 \delta} \frac43 N\right)} = \frac{g_\eta(N; \delta)}{N}.
\end{equation}
Here we've introduced the ``form-factors''
\begin{equation}
g_\epsilon(N; \delta) = \frac{\sqrt{2 \delta}\frac43 N}{\sin \left(\sqrt{2 \delta} \frac43 N\right)}, \quad g_\eta(N; \delta) = \frac{\sqrt{2 \delta}\frac43 N}{\tan \left(\sqrt{2 \delta} \frac43 N\right)}.
\end{equation}

Obviously, the results we've obtained so far in this subsection make sense only for $1 \ll N \ll 1 / \delta$, but this assumption is more than reasonable for our purposes.

So, the leading-order predictions on the inflationary observables of our model are as follows:
\begin{equation}
\label{obs_approx}
n_s \approx 1 - \frac{2 g_\eta (N; \delta)}{N}, \quad r \approx \frac{12 g_\epsilon^2(N; \delta)}{N^2}, \quad A_s \approx \frac{m^2 N^2}{24 \pi^2 M_\mathrm{Pl}^2 g_\epsilon^2(N; \delta)}, \quad \alpha_s \approx -\frac{2 g_\epsilon^2(N; \delta)}{N^2} \approx -\frac{r}{6}.
\end{equation}
One important thing to note is that $g_\eta(N; \delta)$ \textit{decreases} as one increases $\delta$ from $0$ and keeps $N$ fixed. Therefore, this new model can describe $n_s$ values which are larger than ones predicted by the Starobinsky model for the same values of $N$. Also note that, similarly to the Starobinsky model, the $m$ parameter is completely defined by $A_s$, and all the other inflationary observables presented do not depend on it.

\begin{figure}[h!]
\centering
\includegraphics[width=\linewidth]{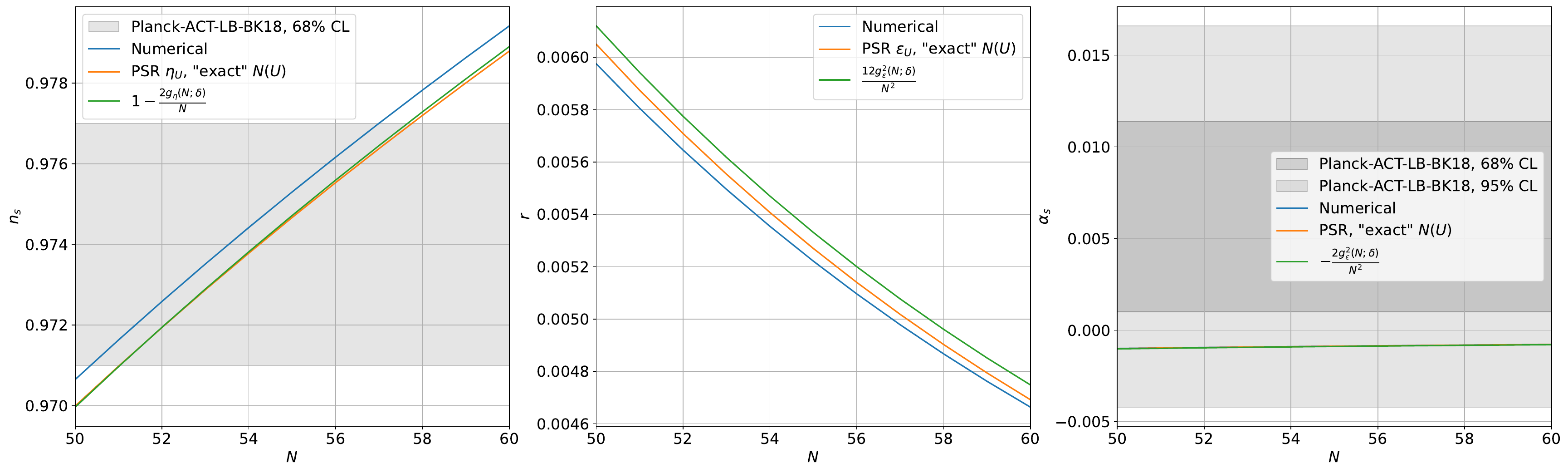}
\caption{$n_s(N)$, $r(N)$, and $\alpha_s(N)$ dependencies for the new model ($\delta = 8\times10^{-5}$). The ``Numerical'' labels refer to the results of numerical integration of dynamical system~(\ref{dyn_syst}); the ``PSR \dots'' labels refer to the results obtained by using the approximate $\epsilon_U$ given by~(\ref{eps_u_psr}), without any additional approximations; the last labels refer to the results obtained by using~(\ref{obs_approx}).}\label{fig:n_s_r_fig}
\end{figure}

Fig.~\ref{fig:n_s_r_fig} shows that the new model can predict values of $n_s$ that conform with the Planck-ACT-LB-BK18 data, given by
\begin{equation}
\label{n_s_data}
n_s = 0.974 \pm 0.003 \quad \text{(68\% CL)}.
\end{equation}

The new model still predicts negative values of $\alpha_s$, while the Planck-ACT-LB-BK18 data implies
\begin{equation}
\alpha_s = 0.006 \pm 0.005 \quad \text{(68\% CL)}.
\end{equation}
It should be noted, though, that the predicted values are still within 95\% CL margins (as seen on Fig.~\ref{fig:n_s_r_fig}).

\begin{figure}[h!]
\centering
\includegraphics[width=0.5\linewidth]{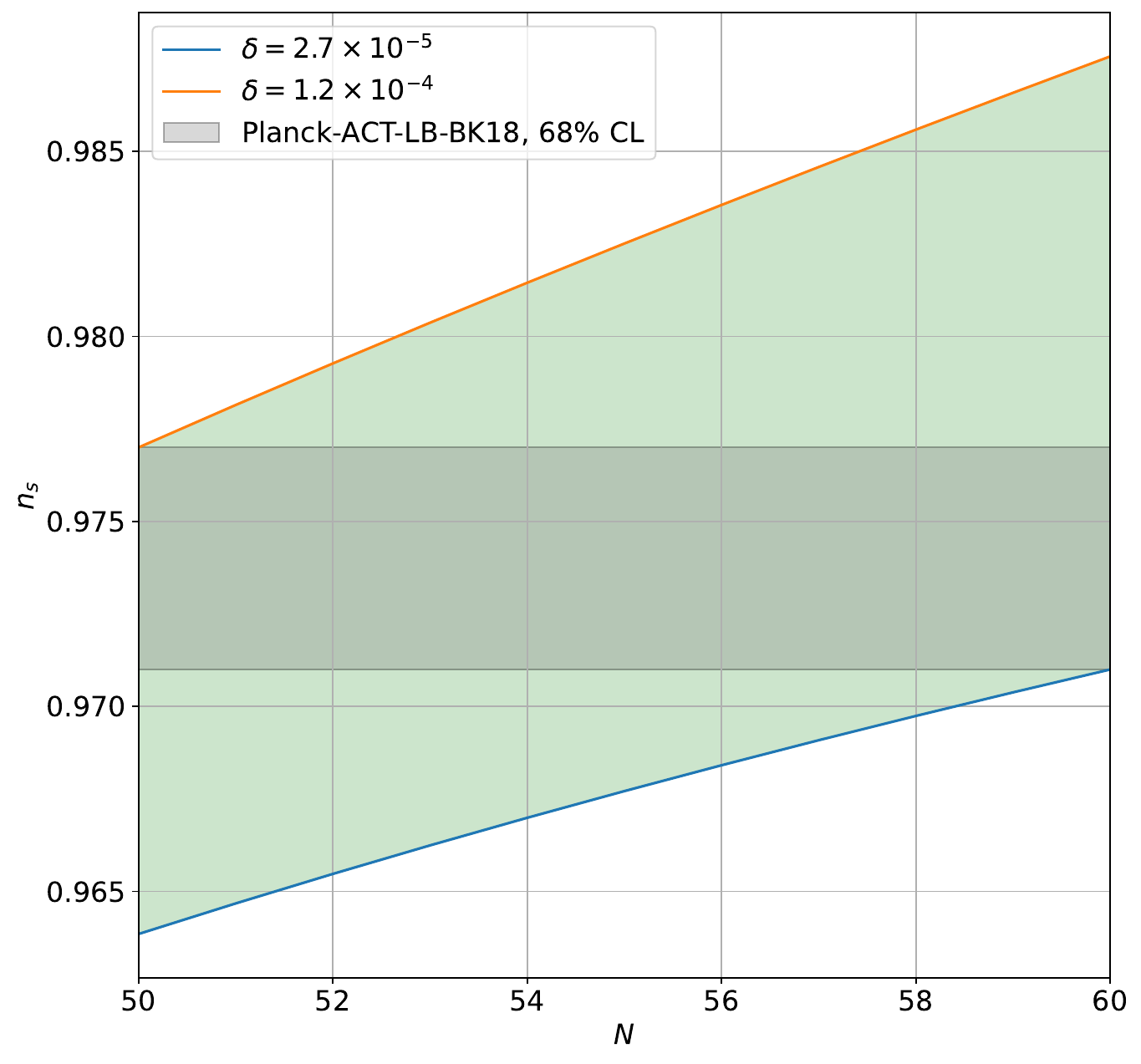}
\caption{$n_s(N)$ dependency for the new model for various $\delta$.}\label{fig:n_s_sweep}
\end{figure}

Now we obtain limits on $\delta$ due to observational constraints~(\ref{n_s_data}), under the condition that $N$ belongs to the interval $(50, 60)$, as is usually assumed in the literature. So, we numerically get the following result~(see also Fig~\ref{fig:n_s_sweep}):
\begin{equation}
\delta \in (2.7 \times 10^{-5},1.2 \times 10^{-4}).
\end{equation}
For all considered values of $\delta$, the values of $u_{end}$ coincide in the first five significant digits: $u_{end} \approx 1.2985$.

\section{Conclusion}

In this paper, we have described the procedure for analyzing inflationary $F(R)$ gravity models in the Jordan frame, without need to refer to the Einstein frame description. We have provided sufficient conditions for the existence of a unique inflationary attractor. The approximate, PSR-like, expressions for the slow-roll parameters and the number of e-folds $N$ were also derived.

The described procedure was applied to two models, the first being the Starobinsky model (which served as a testbench for the described procedure), and the second being a new model, which is a one-parameter deformation of the Starobinsky model. Validity of our PSR-like approximations was confirmed by means of numerical integration of the dynamical system of the model. The new model is able to predict values of the inflationary observables $n_s$, $r$, and $A_s$ that conform with the latest observational data. However, the new model is not able to predict positive values of $\alpha_s$. Nonetheless, in a recent study~\cite{28} the authors show that a certain $F(R)$ model can predict positive values of $\alpha_s$, so inflationary $F(R)$ gravity models are not completely discarded by the requirement of a positive $\alpha_s$.

A possible extension of the research provided in this paper might have to do with analyzing the ultra-slow-roll case. Additionally, it might be promising to generalize the research described to the case of $F(R)$ gravity models with a scalar field or some other form of matter.

\section*{Acknowledgements}
The author would like to express deep gratitude to S.Yu. Vernov for fruitful discussions, comments, and corrections. The author is supported by the ``BASIS'' Foundation grant No. 22-2-2-6-1.

\end{document}